\documentclass{osa-article}

\journal{osajournal}

\usepackage{lineno}

\begin{document}

\title{Correlated twin-photon generation in a silicon nitrite loaded thin film PPLN waveguide}

\author{Antoine Henry,\authormark{1,2}  David Barral,\authormark{1} Isabelle Zaquine,\authormark{2} Andreas Boes,\authormark{3,4,5} Arnan Mitchell,\authormark{3} Nadia Belabas,\authormark{1} and Kamel Bencheikh\authormark{1,*}}

\address{\authormark{1}Centre for Nanosciences and Nanotechnology, CNRS, Universit\'e Paris-Saclay,
UMR 9001,10 Boulevard Thomas Gobert, 91120, Palaiseau, France\\
\authormark{2}LTCI, T\'el\'ecom Paris, Institut Polytechnique de Paris, 19 place Marguerite Perey, 91120, Palaiseau, France\\
\authormark{3}Integrated Photonics and Applications Centre (InPAC), School of Engineering, RMIT University, Melbourne, VIC 3001, Australia\\
\authormark{4}Institute for Photonics and Advanced Sensing (IPAS), University of Adelaide, SA 5005, Australia\\
\authormark{5}School of Electrical and Electronic Engineering, University of Adelaide, SA 5005, Australia}

\email{\authormark{*}kamel.bencheikh@c2n.upsaclay.fr} 



\begin{abstract*}
Photon-pair sources based on thin film lithium niobate on insulator technology have a great potential for integrated optical quantum information processing. We report on such a source of correlated twin-photon pairs generated by spontaneous parametric down conversion in a silicon nitride (SiN) rib loaded thin film periodically poled lithium niobate (LN) waveguide. 
The generated photon pairs have a wavelength centred at 1560\,nm compatible with present telecom infrastructure, a large bandwidth (21\,THz) and a brightness of $\sim 2.5\times 10^5$\,pairs/s/mW/GHz. The photons are correlated and exhibit a cross correlation $g^{(2)}(0)$ of about 8000. Using the Hanbury Brown and Twiss effect, we have also shown heralded single photon emission, achieving an autocorrelation $g^{(2)}_H(0) \simeq 0.04$.    
\end{abstract*}

\section{Introduction}

Bright and wide-band twin-photon sources enable high rate emission of entangled photons and heralded single photons, which are vital resources in various quantum information processing protocols, such as quantum key distribution, optical quantum computing, quantum simulation or quantum metrology\,\cite{Flamini_2018}. 
Twin photons are generally generated by spontaneous parametric down conversion (SPDC), a second order nonlinear interaction process where a single pump photon $\hbar \omega_p$ is annihilated, giving birth to a pair of photons, signal and idler, with respective energies $\hbar \omega_s$ and $\hbar \omega_i$, following the conservation energy relation $\omega_p = \omega_s + \omega_i$. 
Different nonlinear materials have been used to generated twin photons, however lithium niobate (LN) is a reference and competitive material. It has indeed both large second order nonlinear susceptibility ($d_{33} \sim 33$\,pm/V) and electro-optic coefficient ($r_{33} \sim 31$\,pm/V) and a wide transparency window ranging from 0.4\,$\mu$m to 4.5\,$\mu$m.

Lithium niobate integrated devices are conventionally realized in either proton-exchanged or titanium-indiffused waveguides\,\cite{DeMicheli1983, Wooten2000, Parameswaran2002, Bazzan2015}. However, the refraction index difference between the core waveguide and the surrounding lithium niobate is weak for these waveguides, limiting the optical confinement of the guided modes to a few $\mu$m$^2$. Furthermore, the bending radius of such waveguides is in the order of mm, hindering the integration of functional elements on a compact chip. Recently, a new platform based on thin film lithium niobate (TFLN) on insulator has emerged providing much tighter confinement of the optical modes, allowing a more compact circuitry and hence smaller footprint integrated compounds. This new TFLN platform has enabled several breakthrough demonstrations over the last few years, including enhanced nonlinear optical interactions with higher frequency conversion efficiencies\,\cite{Geiss2015, Luo2018, Wang2018, Chen2019, Lu2019, Rao2019b, Zhao2020OE} and efficient and high-speed electro-optic interaction\,\cite{Weigel2018, Desiatov2019, Li2020}.  Whereas second harmonic generation (SHG) has been widely investigated on TFLN, it is only recently that twin photons have been generated and their quantum properties studied on such platform\,\cite{Chen2019, Bradley2019, Elkus2020, Zhao2020PRL}.

Among various waveguide geometries on the TFLN platform, we can distinguish two main fabrication designs. The first one is based on the direct etching of the TFLN to form the waveguides. This is the approach used for example in\,\cite{Chen2019, Bradley2019, Elkus2020, Zhao2020PRL}. However, the etching of lithium niobate can be challenging and the phase matching wavelength is very sensitive to the waveguide dimensions. It offers nevertheless a better confinement thanks to the high refractive index contrast and thus allows for very high second-harmonic (SH) conversion efficiencies, as demonstrated by C. Wang \textit{et al.}\,\cite{Wang2018} in 2018 achieving a SH conversion efficiency of 2600\,\%\,W$^{-1}$\,cm$^{-2}$ and by A. Rao \textit{et al.}\,\cite{Rao2019b} in 2019, measuring an efficiency of 4600\,\%\,W$^{-1}$\,cm$^{-2}$.
The second fabrication strategy consists of depositing a layer of a transparent material with an appropriate refractive index on top of the TFLN and whose processing technology is very mature. The optical mode is confined horizontally by etching this material, resulting in most of the light being confined in the TFLN beneath. This second fabrication strategy is more accessible technologically and despite larger modes, a still-high conversion efficiency can be obtained as it has been reported by A. Boes \textit{et al.} in 2019\,\cite{Boes2019}, achieving about 1160\,\%\,W$^{-1}$\,cm$^{-2}$ SH conversion efficiency.



In this paper, we report to our knowledge the first generation of twin photons via SPDC in periodically poled TFLN waveguides, using the second strategy, depositing and partially etching a thin layer of silicon nitrate (SiN) on top of the lithium niobate thin film. We have successfully generated about $2.5\times 10^5$\,pairs/s/mW/GHz at telecom wavelength with a bandwidth of approximately 170\,nm (21 THz), allowing for wavelength multiplexing and quantum information processing in the frequency domain. Using the strong emission correlation, we have been able to implement heralded single photon emission, achieving a second order heralded autocorrelation function $g^{(2)}_H(0) = 0.04$, very close to the $g^{(2)}(0) = 0$ of an ideal single photon source. 

In the next section, we will give a brief description of the fabrication process of the periodically poled thin film lithium niobate waveguides (PP-TFLN). Then, a classical characterisation through SHG is given in order to determine the appropriate PP-TFLN waveguide for SPDC compatible with our pumping laser source emitting at 780\,nm. Section 3 and 4 are respectively dedicated to the investigation of twin-photon generation properties and to the heralded single photon emission.


\section{PP-TFLN waveguide fabrication}

The periodically poled lithium niobate waveguides were fabricated following similar fabrication steps as described in \cite{Chang2016}. We used commercially available $X$-cut 300-nm thick thin-film lithium niobate wafer from NANOLN, on which we patterned comb-like electrodes with a period of 4.87\,$\mu$m, 4.93\,$\mu$m and 4.98\,$\mu$m, corresponding to the quasi-phase matching periods allowing twin-photon generation at telecom wavelengths when the pump wavelength is set approximately at 800\,nm. We applied high voltage pulses to these electrodes to periodically invert the spontaneous polarisation of the lithium niobate crystal along the $Z$-axis, following the comb pattern of the electrodes, in order to benefit from the highest nonlinear optical coefficient $d_{33}$. Afterwards, we removed the electrodes by wet etching and deposited a 400\,nm thick film of silicon nitride (SiN) \cite{Frigg2019} on the lithium niobate surface. The SiN film was then patterned by photolithography and reactive ion etching, resulting in 2\,$\mu$m-wide SiN stripes at the surface of the lithium niobate thin film. The patterned SiN confines the optical mode horizontally along the $Z$-axis, allowing TE modes for the fundamental (1560\,nm) and second harmonic (780\,nm) wavelengths, with a good modal overlap for efficient nonlinear interaction. Lastly, we coated the sample with a 1\,$\mu$m thick silica layer for protection and prepared the end facets of the waveguides by dicing and polishing.

\section{Classical characterisation of PP-TFLN waveguide}

\begin{figure}[htb]
\centering\includegraphics[width=13cm]{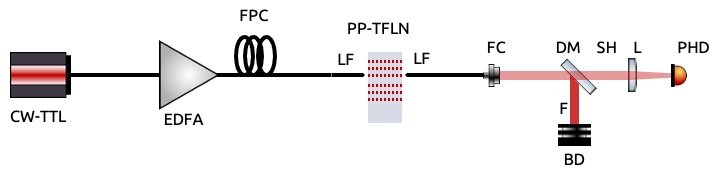}
\caption{Scheme of the experimental setup used to measure SHG from the PP-TFLN waveguide. CW-TTL: continuous-wave telecom tunable laser, EDFA: erbium-doped fiber amplifier, FPC: fibered polarisation controller, FC: fiber collimator, DM: dichroic mirror, BD: beam dump, L: lens, PHD: photodiode, SH: second-harmonic beam, F: fundamental beam}
\label{setup_SHG}
\end{figure}

As a first step, we characterised the PP-TFLN waveguides in order to find the waveguide that is suitable for efficient SPDC when using our pump laser source with a wavelength of 780\,nm. Since SHG is the reverse process of SPDC at degeneracy ($\omega_s = \omega_i$), both phenomena fulfill the same quasi-phase matching condition. Thus, we started by measuring the intensity of the generated second harmonic field when the PP-TFLN waveguides were excited with a laser source tunable around twice the SPDC pump wavelength, using the experimental arrangement depicted in Fig.\,\ref{setup_SHG}. For this purpose, the fundamental field is delivered by a continuous-wave wavelength-tunable laser (TUNICS T100R) emitting about 10\,mW from 1490\,nm to 1620\,nm. To obtain higher powers, we used an erbium-doped fibre amplifier seeded by the laser with a nominal power of 500\,$\mu$W. The amplifier gain is adjusted to achieve  200\,mW at the output which is sufficient for our measurements. The amplified fundamental laser is then coupled into the PP-TFLN waveguides with a lensed fibre from OZ Optics AR-coated at 1550\,nm. To fulfill the phase matching condition, its polarisation is adjusted to excite the TE mode with a fiber polarisation controller (FPC). The generated second harmonic field, also in the TE mode, is collected with a similar lensed fiber (OZ Optics) AR-coated at 780\,nm. On the other side, the lensed fiber is connected to a collimator and the output beam is focused with a 100-mm lens into a free space Si photodiode (Newport 1801-FS), after spectral filtering with a dichroic mirror to remove the fundamental pumping light that might have been collected by the output fibre. In order to improve the detection sensitivity, we modulated the intensity of the fundamental laser by driving its current with a sine-wave function at 80\,kHz. The 30\,\%-depth intensity modulation induced on the fundamental field is transferred to the second harmonic field and detected with the Si photodiode connected to an SRS 100-kHz bandwidth locking-in amplifier, not shown in the figure. The TFLN sample contains 30 periodically-poled waveguides having the same physical length of about 5\,mm and designed in three sets, each with a different poling period: $4.87\,\mu$m, $4.93\,\mu$m and $4.98\,\mu$m. For each set, the poling lengths of the PP-TFLN waveguides are 1.2\,mm, 2.4\,mm, 3.6\,mm or 4.8\,mm. 
The sample is fixed on an aluminum holder and its temperature is controlled and stabilised with a PID-controlled heater. The ensemble is mounted on a 2 axis translation stage which allows to change easily the PPLN waveguide under investigation with minor realignments.


We used this setup to measure the SHG quasi-phase matching wavelength of the waveguides at room temperature. As expected, we found that the second harmonic wavelength is the same within a set of PP-TFLN waveguides and only depends on the poling period. From our measurements, we deduce a phase matching wavelength $\lambda = 1553$\,nm, $\lambda = 1576$\,nm and $\lambda = 1600$\,nm, with an uncertainty of about $\pm$1\,nm, for poling $\Lambda = 4.87\,\mu$m, $\Lambda = 4.93\,\mu$m and $\Lambda = 4.98\,\mu$m respectively. For the last two poling periods, $\Lambda = 4.93\,\mu$m and $\Lambda = 4.98\,\mu$m, the SHG is obtained for fundamental wavelengths larger than $1560$\,nm. Unfortunately, these waveguides are out of reach for SPDC with our pump laser source emitting at 780\,nm or would require cooling of the sample, which we want to avoid to prevent condensation of ambient water vapour on the PP-TFLN waveguides.



\begin{figure}[htb]
\centering\includegraphics[width=13cm]{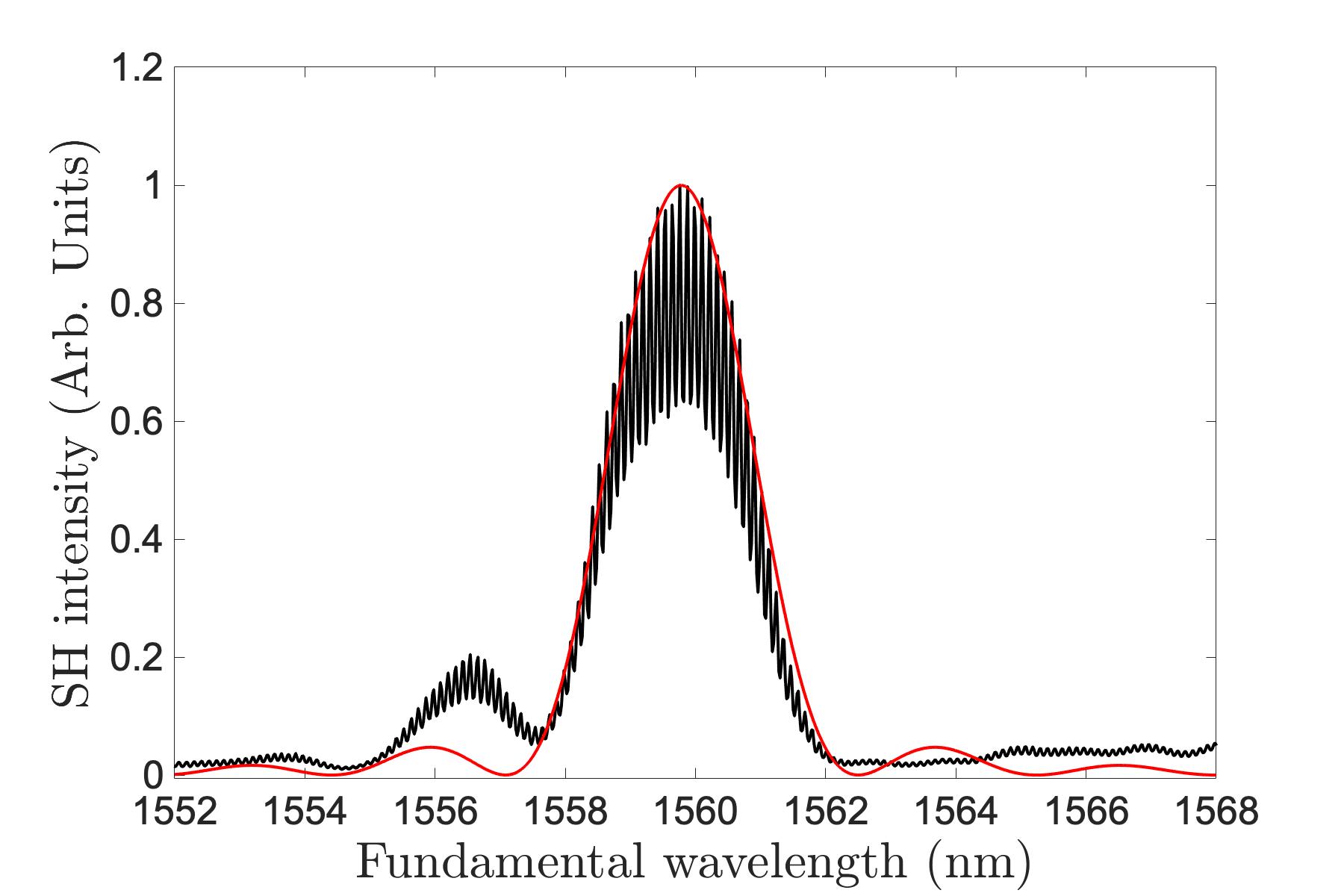}
\caption{Second harmonic generation as a function the fundamental wavelength in the waveguide with a poling period of $\Lambda = 4.87\,\mu$m and poling length $L = 4.8$\,mm. The temperature of the TFLN chip is at 53$^{\circ}$C.}
\label{SHGvswavelength}
\end{figure}

Fortunately, the phase matching wavelength of waveguides with poling period $\Lambda = 4.87\,\mu$m is below $1560$\,nm providing the opportunity to achieve phase matching at $1560$\,nm, by heating the sample up to 53$^{\circ}$C, as shown in Figure\,\ref{SHGvswavelength}. 
The red line is the theoretical phase matching $\mathrm{sinc}^2(x)$ function where $x = (\Delta k -2 \pi/\Lambda) L/2$, $L$ being the interaction length and $\Delta k = k_p-k_s-k_i$, with $k_m = 2\pi n_m/\lambda_m$ being the wavevector of mode $m = p,s,i$.  The red line is  obtained by taking $L=4.8$\,mm and after calculating the effective indices $n_m$ of the TE modes associated with the geometry of the waveguides at the working temperature and working wavelengths. No other extra parameter is used. 
The oscillations in the measured spectra are due to interference effects induced by the reflectivities on the input and output waveguide facets. An oscillation period of about 0.11\,nm is deduced, corresponding to a length of 5.1\,mm, in agreement with the physical waveguides length, which is longer than the poled region. The maximum of the SH field is obtained at 1560\,nm. This indicates that this temperature (53$^{\circ}$C) allows to satisfy the phase-matching condition when pumping the PP-TFLN waveguide at 780\,nm to generate by SPDC twin photons having wavelengths around 1560\,nm. 

\section{SPDC and quantum characterisation of PP-TFLN waveguide}

\begin{figure}[htb]
\centering\includegraphics[width=13cm]{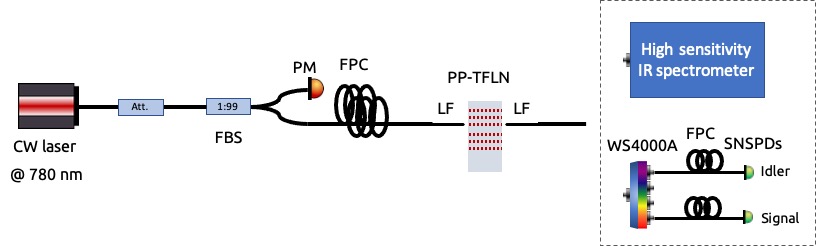}
\caption{Scheme of the experimental setup used to characterize SPDC using either a high sensitivity infrared (IR) spectrometer or single photon detectors (SNSPDs) with a waveshaper to separate signal from idler photons. Att: fibred optical attenuator, FBS: fibred beamsplitter, PM: powermeter, WS4000A: waveshaper 400S, SNSPDs: superconducting nanowire single-ophoton detectors.}
\label{fig_spdc}
\end{figure}

In order to implement spontaneous parametric down conversion and to measure the twin photon emission, we use the experimental setup depicted in Fig.\,\ref{fig_spdc}. The source is a high-power cw fiber-coupled laser emitting up to 1 W at 780\,nm. For our demonstrations we set the laser output power at 200\,mW, which is more than sufficient. The power is tuned with a fibre variable attenuator. The monitoring of the pump power is achieved by sending the  1\,$\%$ output port of a 1:99 fiber coupler to a powermeter (NEWPORT). On the 99\,\% port, two cascaded 50:50 fibre coupled beamsplitters (not shown in the figure) are used to reduce the power and a fibre polarisation controller is used to set the pump polarisation, before coupling to the PP-TFLN waveguide with a lensed fibre (AR-coated at 780\,nm). 

As a first step, we estimated the coupling efficiency into the waveguide. This is done by measuring the power at the output of the lensed fibre using a free space optical powermeter. We then use this fibre to couple the 780\,nm wavelength beam into the PP-TFLN waveguide. A similar lensed fibre is used to collect the guided pump laser. Using the same powermeter we measured the transmitted pump and inferred the coupling efficiency to be $\sim$\,5\,$\%$ at 780\,nm, where we have made a reasonable assumption that the input and output couplings are the same. Following the same procedure, we have deduced a coupling efficiency of about $\sim\,29\,\%$ at 1560\,nm by replacing the 780\,nm lensed fibers with 1560\,nm lensed fibres. We have to emphasise here that no particular design has been done in order to improve the coupling efficiencies. 

\begin{figure}[htb]
\centering\includegraphics[width=13cm]{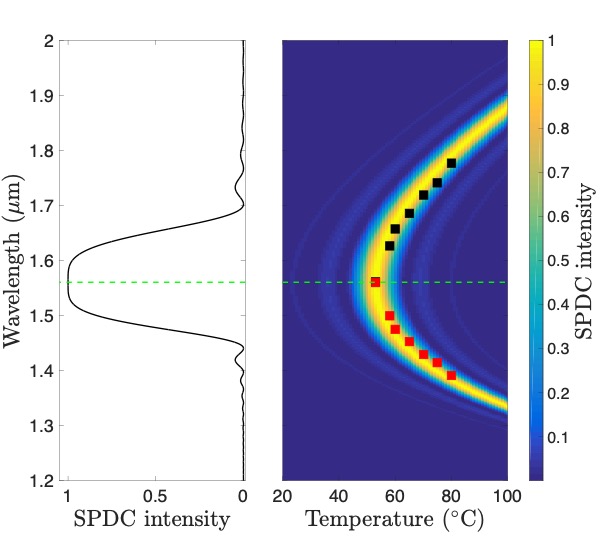}
\caption{Left: Spectral SPDC bandwidth at degeneracy reached for a phase matching temperature of 53$^{\circ}$C. Right: signal and idler wavelengths for different phase matching temperatures. The red squares indicate the measured signal wavelengths and black squares the deduced idler wavelengths. The background map is the calculated phase matching using the poling period  $\Lambda = 4.87\,\mu$m and interaction length of 4.8\,mm. The green dashed horizontal line highlights the degeneracy point.}
\label{SPDCvsT}
\end{figure}

Having estimated the coupling efficiencies, we set back the lensed fibre AR coated at 780\,nm to pump the PP-TFLN waveguides and kept the one AR coated at 1560\,nm to collect twin photons. A fine determination of the SPDC phase matching temperature is obtained by recording with a high-sensitivity infrared spectrometer the wavelengths of the twin photons generated by the SPDC as a function of the temperature. As the spectrometer is based on a 1D InGaAs array with 532 pixels, we have only been able to measure the SPDC for the signal photons, because the spectral sensitivity of the spectrometer drops drastically above 1.6\,$\mu$m at the low ($\sim$70\,K) working temperature of the spectrometer. The measured wavelengths are reported as red squares in Fig.\,\ref{SPDCvsT} as a function of the temperature of the TFLN chip. The black squares are the wavelengths of the corresponding idler photons deduced from the energy conservation relation $1/\lambda_p=1/\lambda_s+\lambda_i$. The background map is the theoretical  $\mathrm{sinc}^2(x)$ function where $x = \Delta k - 2 \pi/\Lambda$ and $L=4.8$\,mm is the interaction length. It is calculated using Lumerical finite-difference eigenmode solver to simulate the effective mode indices for the full range of wavelengths and temperatures ranging from 20$^{\circ}$C to 120$^{\circ}$C for our geometrical design of the PP-TFLN waveguides. As in the case of SHG, we have a very good agreement between the measured phase matching wavelengths and predicted ones, confirming the quality of the PP-TFLN waveguides fabrication.

When setting the temperature to 53$^{\circ}$C, the signal and idler are degenerate. However due to the short interaction length, the bandwidth of the twin photons is very large, extending over $\sim$\,21\,THz. This large bandwidth allows easy addressing of both degenerate and non-degenerate configurations using a programmable optical filter WaveShaper 4000S. As depicted in Fig.\,\ref{fig_spdc}, we connect the lensed fibre collecting the twin photons to the programmable filter. In the non-degenerate case, we program the filter such as the first output has a 100\,GHz bandwidth flat-top filter centered at $f_0+400$\,GHz away from the degeneracy at 1560\,nm ($f_0$ = 192.113\,THz), which corresponds to the signal photons. A second identical band-pass filter is programmed on the second output of the programmable filter, but centered at $f_0-400$\,GHz from degeneracy, which corresponds to the idler photons. Both outputs are then directed to superconducting nanowire single-photon detectors (SNSPDs) from Single Quantum to count signal and idler photons and to monitor the coincidences using Time Tagger Ultra from Swabian Instruments. Both detectors have a quantum efficiency of 70\,\%, a dead time of 20\,ns and a jitter of about 30\,ps. 

\begin{figure}[htb]
\centering\includegraphics[width=13cm]{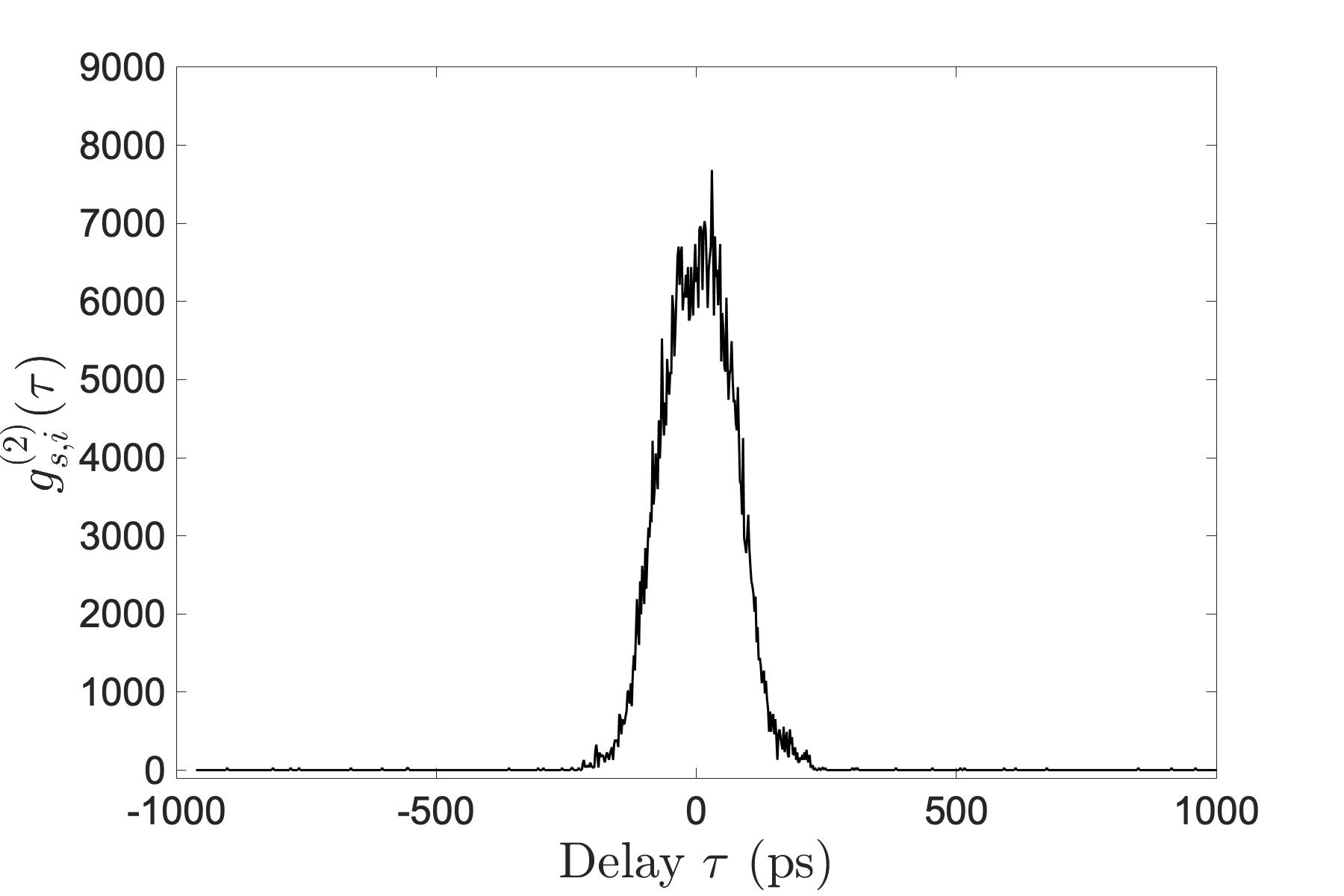}
\caption{Signal and idler coincidence function $g^{(2)}(\tau)$ for different delays between the signal and idler photons.}
\label{g2vsT}
\end{figure}

Next, we measured a typical normalised temporal histogram of the coincidences between signal and idler photons obtained with time bins of 10\,ps, which is shown in Fig.\,\ref{g2vsT}. The central peak at zero indicates that the signal and idler photons are indeed twins created simultaneously in the PP-TFLN waveguide, arriving at the same time to the detectors within a time resolution limited by the detector jitters, the resolution of the time tagger and by the coherence of the photons fixed by the bandwidth of the filters. 

\begin{figure}[htb]
\centering\includegraphics[width = 14cm]{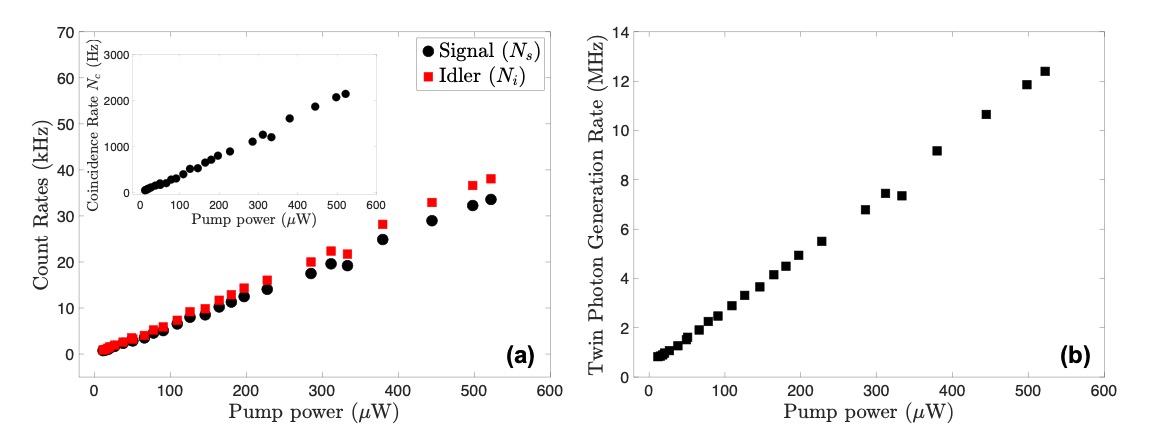}
\caption{(a): Count rates of the signal $N_s$ and idler $N_i$ photons and the coincidence rate $N_c$. (b): Evolution of the internal twin-photon generation rate for different pumping powers. The signal and idler photons have a bandwidth of 100\,GHz.}
\label{twinratevspower}
\end{figure}

In order to determine the internal twin-photon generation rate $N_t$, we have measured the signal $N_s$ and idler $N_i$ photons rates as well as the coincidence rate $N_c$ in a time window ($\sim$600\,ps) larger than the width of the central peak shown in Fig.\,\ref{g2vsT} for different pump powers coupled into the PP-TFLN waveguide. In our case, the twin rate is simply given by $N_t = (N_s \times N_i) / N_c$ and thus does not require as a prerequisite the determination of the different optical losses of our experiment setup, including the detectors quantum efficiencies. Figure\,\ref{twinratevspower}(a) shows the three rates measured for different pump powers and in Fig.\,\ref{twinratevspower}(b) the corresponding internal twin-photon generation rates. A linear fit of the results in Fig.\,\ref{twinratevspower} (b), allows to extract a twin generation rate coefficient of 23\,MHz/mW. Moreover, as the signal and idler photons go through identical band-pass filters of 100\,GHz, we can deduce the brightness of our source to be $2.3\times 10^5$\,pairs/s/mW/GHz, a value close to the brightness of $4.6\times 10^5$\,pairs/s/mW/GHz reported by J. Zhao \emph{et al.} \,\cite{Zhao2020PRL} in PP-TFLN fabricated by etched lithium niobate.

Knowing the internal twin-photon generation rate, we can deduce the overall optical losses on each channel. Indeed, the signal and idler measured rates are given by $N_{k = s,i} = \eta_k \eta_D N_t$, where $\eta_D = (70 \pm 2)\,\%$ is the quantum efficiency of the detectors and $\eta_k$ is the transmission efficiency taken into account all optical losses from the waveguide until the output of the final optical fibres connected to the detectors. From our results we have been able to deduce $\eta_s = 8\,\%$ and $\eta_i = 9\,\%$ with an uncertainty of about $\pm 2\,\%$ deduced statistically, repeating the measurement several times. These efficiencies are coherent with the $29\,\%$ coupling efficiency at 1560\,nm into the lensed fibre and taking into account the $\sim$\,6\,dB losses due to the waveshaper and the losses of $\sim$\,0.5\,dB at each fibre connection. 

\begin{figure}[htb]
\centering\includegraphics[width=12cm]{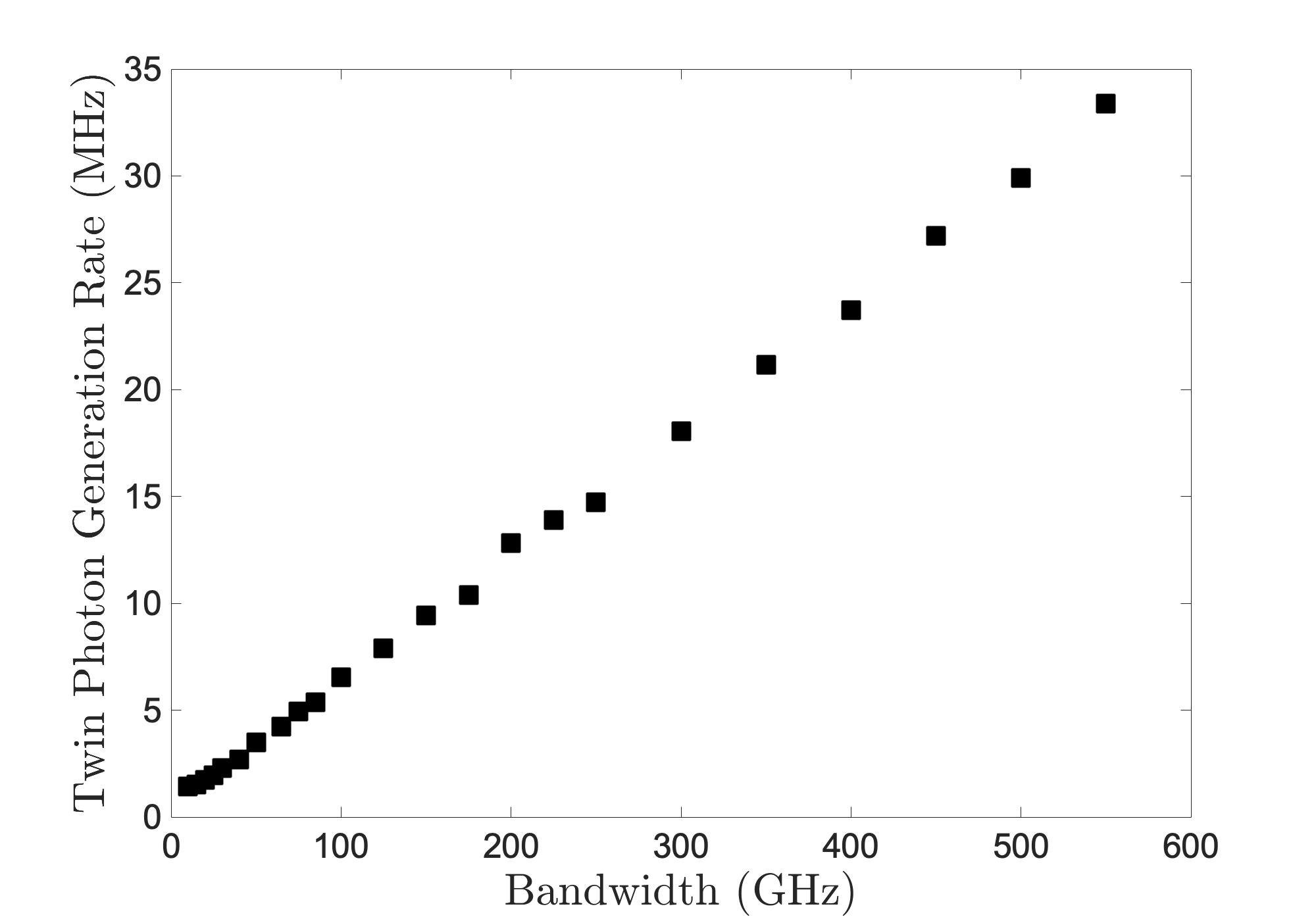}
\caption{Evolution of the twin-photon generation rate for different Signal and Idler photons bandwidths. The pump power is 180\,$\mu$W.}
\label{twinratevsbdw}
\end{figure}

We  also independently determined the SPDC brightness, namely by fixing the pump power (180\,$\mu$W) and tuning the bandwidth of the flat-top filters in the signal and idler channels. In order to reach large bandwidths and avoid any overlap between the signal and idler photons, we shifted the central frequencies of the filters by $f_0\pm 800$\,GHz away from the degeneracy. As shown in Fig.\,\ref{twinratevsbdw}, the measured internal twin-photon generation rate grows linearly with the filter bandwidth. From the linear fit we deduce a slope of 0.058\,MHz/GHz, which corresponds to a brightness of $\sim 3\times 10^5$\,pairs/s/mW/GHz, close to the brightness we have found above.  

\begin{figure}[htb]
\centering\includegraphics[width=12cm]{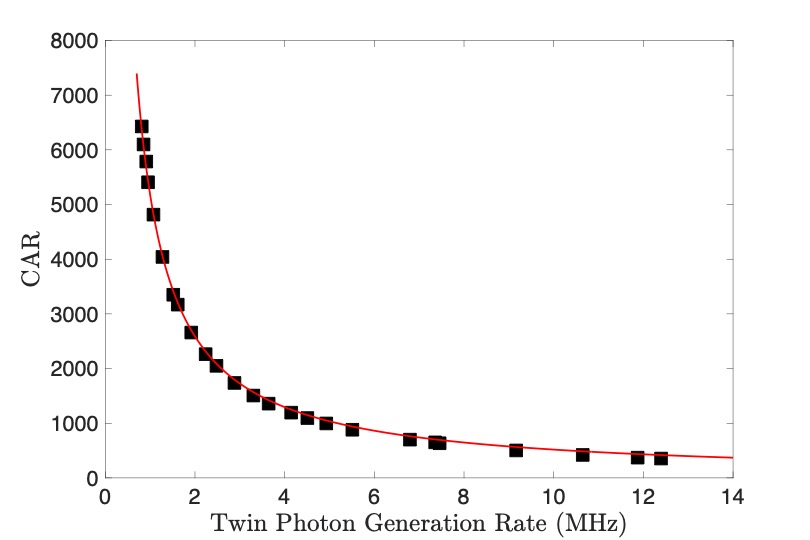}
\caption{Coincidence-to-Accidental Ratio as a function of the twin-photon generation rate $N_t$ measured using flat frequency filters of 100\,GHz for the signal and idler.}
\label{Fig-CAR}
\end{figure}

After having determined the brightness of our source, the generated twin photons are further investigated by measuring the cross correlation between the signal and idler photons, using the second order correlation function\,\cite{Loudon}
\begin{equation}
g^{(2)}_{i,s}(\tau) = \frac{\langle \hat{a}^{\dagger}_s (t) \hat{a}^{\dagger}_i(t+\tau) \hat{a}_i(t+\tau) \hat{a}_s(t) \rangle}{\langle \hat{a}^{\dagger}_s \hat{a}_s \rangle \langle \hat{a}^{\dagger}_i \hat{a}_i \rangle }.
\end{equation}
Figure\,\ref{g2vsT} is a typical signal-idler cross correlation function showing clearly the bunching of the twin photons. For uncorrelated photons, such as those emitted by a coherent laser, the cross correlation $g^{(2)}$ is unity, and is independent of the delay between the arrival time of the photons on the detectors. The cross correlation $g^{(2)}_{s,i}(\tau)$ is measured by sending the signal and idler to the SNSPD detectors and collecting the time tags of the generated electrical pulses with the time tagger. The true coincidences-to-accidental ratio (CAR) is obtained by correcting  $\max(g^{(2)}_{s,i}(\tau = 0))$ from the accidental coincidences for which $g^{(2)} = 1$.  Figure\,\ref{Fig-CAR} represents the CAR for different twin-photon rates, hence different pump powers,  for a 100-GHz bandwidth filters on the signal and idler photons. As expected the CAR is proportional to the inverse of the generation rate, following the theoretical analysis given in Ref.\,\cite{Clausen2014}, with a high value at low rates indicating a high level of photon bunching.  

SPDC is a resource for producing bunched twin photons as shown in Fig.\,\ref{Fig-CAR}. This is an extraordinary resource that is harnessed to produce heralded single photon sources. Indeed, in the weak interaction regime, CAR\,$\gg 1$, the wave function of the generated twin photons can be approximated by 
\begin{equation}
| \psi \rangle  \simeq   |0,0 \rangle + \epsilon |1_{\omega_s},1_{\omega_i}\rangle,
\label{twin_state}
\end{equation}
where the coefficient $\epsilon$ accounts for the nonlinear interaction efficiency and the phase matching bandwidth. The bi-photon state Eq.\,(\ref{twin_state}) indicates that a single pair of signal and idler photons is generated with a probability $|\epsilon|^2/(1+|\epsilon|^2) \ll 1$. By detecting for example the signal photon emitted at frequency $\omega_s$, we project the state into $| \psi \rangle  = |1_{\omega_i}\rangle$ with one idler photon at frequency $\omega_i = \omega_p - \omega_s$. Implementing such heralded single photon sources on the compact and very efficient TFLNOI platform is a real asset for integration with telecom C-band infrastructure. Another asset of such a source is its large bandwidth enabling for wavelength multiplexing, broadband tunability of the emitted single photon or fast repetition rates when SPCD is excited with short pulses. In this work, the tunability is about 100\,nm for the PP-TFLN waveguide with $\Lambda = 4.87\,\mu$m at $T=53\,^{\circ}$C, as shown in Fig.\,\ref{SPDCvsT}. It can be extended to up to 300\,nm by changing the working temperature or shortening the interaction length.


In order to demonstrate that the idler photon is indeed a Fock state $|n=1\rangle$, we measured its autocorrelation function which should be $g^{(2)}(0) = 0$. This is achieved through a Hanbury-Brown and Twiss experiment\,\cite{Hanbury1956} by connecting a 50:50 fibre beamsplitter at the idler output of the waveshaper and sending the two beamsplitter outputs, called idler$_1$ and idler$_2$, to the SNSPD detectors having a 70\,\% quantum efficiency. The signal photon is detected with a third SNSPD detector having a quantum efficiency of 30\,\%. With the time tagger card, we measured simultaneously the triple coincidences signal-idler$_1$-idler$_2$ and all bi-coincidences signal-idler$_1$, signal-idler$_2$ and idler$_1$-idler$_2$ in a time window of about 600\,ps, slightly larger than the cross-correlation temporal width. The count rate on each channel is also recorded. To avoid long term drifts and have fair enough signal-to-noise ratio, we collected the different coincidences and rates over one minute. Finally we deduce the $g^{(2)}(0)$ of the idler photon heralded by the detection of the signal photon\,\cite{Grangier1986, Beck2007}, which is given by 
\begin{equation}
g^{(2)}_H(0)= \frac{C_{s,i1,i2} \,\, N_s}{C_{s,i1}\,\,C_{s,i2}},
\end{equation}
where $C_{s,i1,i2}$ is the triple coincidence rate, $N_s$ the count rate of the signal photons, and $C_{s,i1}$ ($C_{s,i2}$) is the coincidence rate between the signal and idler$_1$ (signal and idler$_2$). 

\begin{figure}[htb]
\centering\includegraphics[width=12cm]{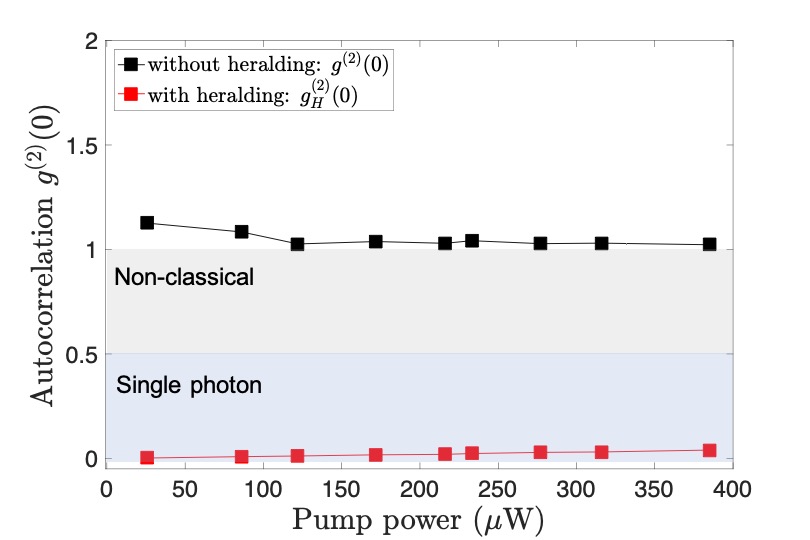}
\caption{Autocorrelation $g^{(2)}(0)$ of the idler photon as a function of the pump power.}
\label{Fig-g2idler}
\end{figure}

Figure\,\ref{Fig-g2idler} shows the  autocorrelation function $g^{(2)}(0)$ of the idler for different pump powers, measured when heralding by the detection of the signal photon (red) and without heralding (black).
The light blue rectangle indicates the area in which the source is considered to be in the single photon state regime, whereas the  light gray rectangle highlights the $g^{(2)}$ of Fock states $| n\rangle$ with more than one photon. In Fig.\,\ref{Fig-g2idler}, $g^{(2)}_H (0) \simeq 0 $ indicates clearly single photon state emission. At a pump power of 385\,$\mu$W, we measured $g^{(2)}_H (0) \simeq 0.04$, corresponding to a triple coincidence rate $C_{s,i1,i2} = 14$\,Hz, a signal and idler detection rates of about 428\,kHz, and 987\,kHz respectively and a coincidence rate between the signal and idler around 48\,kHz. The factor 2 between the signal and idler detection rates is due to the unbalanced SNSPDs quantum efficiencies used for this demonstration. 

Figure.\,\ref{Fig-g2idler}  also exhibits a slight increase of $g^{(2)}_H (0)$ as the pump power is increased. This is due to the fact that the single photon character of the source as described by Eq.\,(\ref{twin_state}) depends on the strength of $\epsilon$ which is pump-power dependent or equivalently dependent on the twin-photon generation rate $N_t$. As the pump power increases, the probability of generating more than a single pair within the measurement time window cannot be neglected anymore. This gives rise to triple coincidence counts during the measurements and thus to a higher value of $g^{(2)}_H (0)$. 

When measuring the $g^{(2)}(0)$ without taking into account the detection of the signal photons, we retrieve $g^{(2)}(0)\gtrsim 1$, indicating that our source is emitting twin photons in a multimode regime with a statistic close to the Poisson statistics and thus exhibits a coherent state behaviour. This is due to the fact that the pumping source has a very long coherence time (kHz bandwidth), whereas the photons are filtered with 100\,GHz filters corresponding to short coherence time, of the order of a picosecond. This indicates that we are selecting much more than one signal or idler modes with the band-pass filters. The slight higher values of $g^{(2)}(0)$ for pump powers $<150\,\mu$W is attributed to statistical errors as the number of coincidence counts is very low. A longer measurement time is required to infer a $g^{(2)}(0)$ value with good uncertainty. In the case of an ideal single mode detection, $g^{(2)}(0)= 2$, a signature of thermal statistics characterising twin photons emission\,\cite{Loudon}. This situation could be reached by further narrowing for example the signal and idler filters down to the MHz level, close the linewidth of the pumping source. However, the count rates, and particularly the triple coincidence rate, will be very low. It is then preferable to pulse the excitation source in order to increase its spectral bandwidth to make it compatible with the twin photons detection bandwidth which enables access to high rates.

\section{Conclusion}
Twin photon sources are important building blocks for quantum information processing. We have reported here on an integrated source on the thin film LN platform. Unlike traditional sources based on LN thin film where the LN is etched, here the waveguide is defined by first depositing a thin layer of SiN on top LN, followed by a shallow etching of SiN using well known and mastered nano-fabrication techniques. Low loss, large transparency bandwidth SiN is a also a promising platform when combined with efficient TFLNOI for optical quantum technologies. A brightness of about $\sim 10^5$ pairs/s/mW/GHz is achieved, allowing the demonstration a single photon emission. These results pave the way to on-chip quantum applications based on twin photons and advantage of their quantum correlations.

We acknowledge the support of the international Associated Laboratory in Photonics between France and Australia (ALPhFA$+$). This work was supported by the Australian Research Council’s Discovery Project scheme (DP190102773). It has also been supported by Region Ile-de-France in the framework of DIM SIRTEQ and by the Agence Nationale de la Recherche through Project INCQA.



\bibliography{source_file.bib}






\end{document}